\documentclass{aa}
\usepackage{graphics,astron}
\title{Simultaneous Multifrequency Observations of the BL Lac \object{MS 0205.7+3509}}
\author{D.\,Watson\inst{1} \and L.\,Hanlon\inst{1} \and 
	B.\,McBreen\inst{1} \and N.\,Smith\inst{2} \and
	M.\,Tashiro\inst{3} \and A.R.\,Foley\inst{4} \and
	L.\,Metcalfe\inst{5} \and V.\,Beckmann\inst{6} \and
	S.F.\,S\'anchez\inst{7} \and H.\,Ter\"{a}sranta\inst{8} }
\authorrunning{Watson et al.}
\titlerunning{Multifrequency Observations of the BL Lac \object{MS 0205.7+3509}}
\institute{Department of Experimental Physics, University College Dublin, Belfield, Dublin 4, Ireland
        \and Dept. of Applied Physics and Instrumentation, Cork Institute of Technology, Cork, Ireland
        \and Department of Physics, University of Tokyo, Hongo, Bunkyo-ku, Tokyo, 113-0033, Japan
        \and Netherlands Foundation for Research in Astronomy, Dwingeloo, The Netherlands
        \and ISO Science Operations Centre, Villafranca del Castillo, Madrid, Spain
        \and Hamburger Sternwarte, Gojenbergsweg 112, D-21029 Hamburg, Germany
        \and IFCA-Dept.\,de F\'{\i}sica Moderna, CSIC-Univ.\,de Cantabria, Santander, Spain
        \and Mets\"{a}hovi Radio Research Station, Kylm\"{a}l\"{a}, Finland}
\offprints{dwatson@bermuda.ucd.ie}
\date{Received / Accepted }
\thesaurus{3(11.02.2 \object{MS 0205.7+3509};  12.07.1;  13.25.2)}

\begin{document}
\maketitle

%----------------------------ABSTRACT----------------------------------

\begin{abstract}
	Radio and optical observations of the possible microlensed BL Lac
	source \object{MS 0205.7+3509} were obtained simultaneously with ASCA
	x--ray measurements in February 1997.  A single power law model,
	with a photon index of 2.61, is an adequate fit to the ASCA
	data, once hydrogenic absorption in excess of the Galactic value is
	permitted, confirming a previous ROSAT measurement.
	On the basis of our simultaneous data we have determined
	\object{MS 0205.7+3509} to be a typical x--ray selected BL Lac, with
	$\alpha_{xox}=-0.82$.
        There is no
	indication of an inverse Compton (IC) component in the ASCA spectrum
	up to 10\,keV.  No evidence for variability on hour-long timescales
	is present in either the x--ray or the optical data.  We discuss
	these results in the context of a gravitational microlensing
	scenario for \object{MS 0205.7+3509}.

\end{abstract}

%-------------------------INTRODUCTION----------------------------------

\section{Introduction}
The primary distinguishing characteristics of BL Lacs, such as their high
and variable polarisation, extreme variability at optical through radio
wavelengths, apparent superluminal motion and their lack of prominent line
emission (equivalent width $<5\AA$), may be explained by relativistic
beaming effects in jets which are oriented close to the line of sight
\cite{1995PASP..107..803U}.  Significant differences between x--ray selected
BL Lacs (XBLs), discovered in x--ray surveys and radio-selected BL Lacs
(RBLs), discovered in radio surveys, may be attributable to differences in
jet direction relative to the observer and in the width of the cone into
which the emitted radiation is beamed
\cite{1991ApJS...76..813S,1991ApJ...374..431S,1989ApJ...340..181G}.  A
smaller angle of the jet to the line of sight is inferred for RBLs on the
basis of the greater optical variability and higher and more variable
polarisation exhibited by these sources compared to XBLs \cite{jse93,jse94}. 
In this orientation model for BL Lacs there should be a significant
population of sources with properties intermediate between those of XBLs and
RBLs. The recent identification of such a population in the ROSAT All-Sky
Survey/Green Bank (RGB) sample of BL Lacs provides further evidence in
favour of the orientation hypothesis \cite{Laurent-Muehleisen:1998}. 
However, orientation effects alone might not be able to account for the
different spectral energy distributions of RBLs and XBLs, which may require
differences in intrinsic physical parameters of the jet
\cite{1996ApJ...463..444S}.

In the context of unified schemes for Active Galactic Nuclei (AGN), BL Lacs
are expected to reside in luminous elliptical galaxies and should be centred
with respect to these host galaxies
\cite{1991MNRAS.252..482A,1996ApJS..103..109W,1995ApJ...454...55S}. However
several BL Lacs have been identified (e.g. \object{MS 0205.7+3509},
\object{AO 0235+164}, \object{PKS 0537-441} and \object{W1 0846+561}
\cite{1993AAS...182.0410W,1995ApJ...454...55S}) which are not centred with
respect to their hosts or their surrounding nebulosities. Furthermore, in a
few cases (e.g. PKS\,1413+135, OQ\,530) the underlying host galaxy appears
to be spiral rather than elliptical \cite{mmac94}.  Excess soft x-ray
absorption and/or an unusual morphology (i.e. spiral host galaxy or
decentering of the nucleus) in these BL Lacs may indicate the presence of a
foreground galaxy and suggests that microlensing effects may be important to
explain all of the observed properties of these sources
\cite{1985Natur.318..446O,1993ApJ...410...39G,Stickel:90}. We note
that HST imaging of PKS\,1413+135 has been used to rule out the presence of
gravitational lensing effects in this source \cite{mmac94}.

\object{MS 0205.7+3509} is a candidate for microlensing due to the possible
spiral nature of its `host' galaxy and the excess soft x--ray absorption
which has been determined from ROSAT PSPC observations
\cite{1995ApJ...454...55S}.  A redshift of $z=0.318$ has
been determined from a MgII absorption feature in the optical spectrum of
\object{MS 0205.7+3509} \cite{1991ApJ...380...49M}. However, the almost
featureless optical spectrum makes the redshift uncertain and more recent
attempts to confirm it have not been successful
\cite{Stocke1998}. \object{MS 0205.7+3509} has been classified as an XBL in
the complete BL Lac sample of the {\em Einstein} Extended Medium Sensitivity
Survey (EMSS), where it was first identified
\cite{1991ApJ...380...49M,1990ApJS...72..567G,1996ApJ...456..451P}.

In this paper, we present the results of simultaneous x--ray, optical and
radio observations of \object{MS 0205.7+3509}, interpret the data and
discuss the gravitational lensing hypothesis.

%------------------------------OBSERVATIONS & REDUCTION--------------------------------

\section{Observations and Data Reduction}

Observations of \object{MS 0205.7+3509} were made with the ASCA satellite
\cite{1994PASJ...46L..37T} between 00:34 and 18:58\,UT on 6 February 1997. 
There are four instruments on board ASCA, two Gas Imaging Spectrometers
(GIS-S2 and GIS-S3) and two Solid-state Imaging Spectrometers (SIS-S0 and
SIS-S1). The instruments have a well-calibrated energy range of
0.7--10.0\,keV for the GIS and 0.5--10.0\,keV for the SIS and only data in
these energy ranges were used for spectral fitting. The total observation
time per instrument was approximately 35\,ks.  The data were recorded in
FAINT data mode and converted to BRIGHT-2 mode \cite{1993ExA.....4....1I}
having corrected the SIS data for the Residual Dark Distribution (RDD)
effect with the FTOOLS script `correctrdd'.

Data from the SIS and GIS instruments were screened in a standard way, and
were then reduced using the \emph{Ftools} applications. An extraction region
was defined around the source, giving an aperture of 4\arcmin\, for the SIS
detectors and 6\arcmin\, for the GIS. There are no other sources in the
extraction region down to a (6\,cm) flux limit of 50\,$\mu$Jy (see
Fig.~\ref{radio_map}).  This spatial region was used to extract the source
plus background counts.  An estimate of the background was derived by taking
an annulus around the extraction aperture and extracting counts from this
region. An alternative background estimate was derived from separate
dark-sky observations made by ASCA and results obtained using both types of
background estimates were found to be consistent. The annulus background was
subtracted from the extracted source plus background counts to obtain a
source spectrum.  The resulting spectra were rebinned to have a minimum of
10 counts per bin in order to ensure the validity of the Gaussian
approximation.

The flux densities used in $\log(S_{x}/S_{r})$ and in calculating $\alpha_{ro}$,
$\alpha_{rx}$ and $\alpha_{ox}$ have been K-corrected by multiplying observed
flux densities by $(1+z)^{\alpha-1}$, where $\alpha$ is the power-law spectral
index in the appropriate energy band \cite{1996ApJ...463..444S}.  $\alpha_{r}$
is derived from a linear interpolation of flux densities given by Stocke et al.
\cite*{1995ApJ...454...55S};  $\alpha_{o}$ is derived from the optical spectrum
of \object{MS 0205.7+3509} \cite{1991ApJ...380...49M}; $\alpha_{x}$ is taken
from these observations.

\subsection{Spectral Fitting}

The combined binned spectra were fit simultaneously using the
Levenberg-Marquardt algorithm in \emph{XSPEC}.  The data were fit to: (a) a
model of a power law, with an absorber as defined by Morrison and McCammon
\cite*{1983ApJ...270..119M} fixed at the Galactic value
\cite{1992ApJS...79...77S}, (b) a power law model, with a
local absorber fixed at the Galactic value and a second absorber at a
redshift $z=0.318$ and (c) broken power law with a local absorber fixed at
the Galactic value.

The fits to models (a) and (b) were compared using the F-test
\cite{1992drea.book.....B}. Fits to models (b) and (c) were
also compared in this way.

Data from the four ASCA instruments (the two SIS and the two GIS detectors
are referred to from now on as S0, S1 and S2, S3 respectively) were compared
for consistency, using the best fit model results (see 
Table~\ref{instrumentresults}) and were found to be compatible.

\subsection{Optical and Radio Observations}

Optical Johnson V-band observations were carried out under photometric conditions
with the 1.23\,m telescope at Calar Alto Observatory and with the 1\,m JKT at La
Palma on the nights of 5 and 6 February 1997 with typical exposure times of
between 1200\,s and 1800\,s.
Standard CCD aperture photometry techniques were used to reduce the data.
The relative uncertainty (0.2 mag.) in the optical magnitude was brought below
the absolute calibration uncertainty, by reducing the size of the aperture from
which the flux was extracted to achieve optimal signal to noise.  The flux
in this optimal aperture could then be used to compare different observations
of the object, on the same night, for evidence of relative variability.
\object{MS 0205.7+3509} was also monitored daily from
February 4--7 1997 at 4986.99\,MHz (6\,cm) with the Westerbork Synthesis
Radio Telescope for integration times of $\sim 12$ hours.

%------------------------------RESULTS------------------------------------

\section{Results}

\begin{figure}
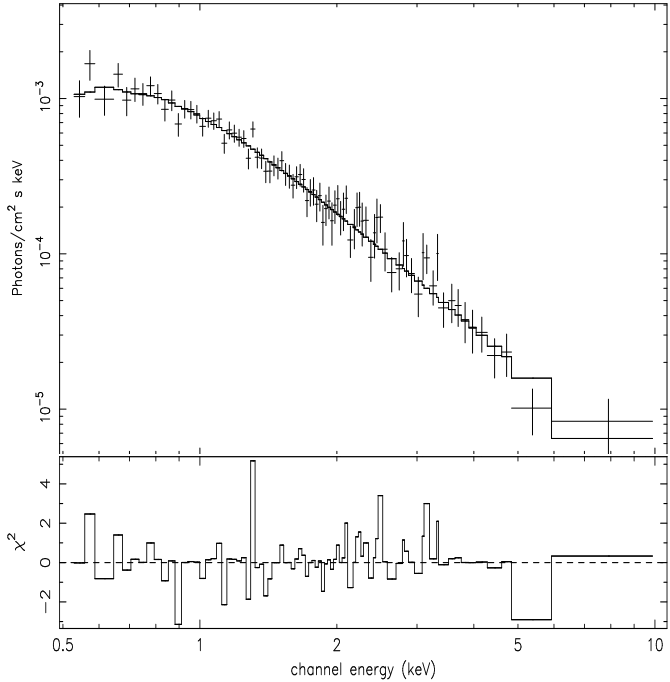

	\rotatebox{-90}{\resizebox{!}{\columnwidth}{\includegraphics*{h1235.f1a}}}
	\rotatebox{-90}{\resizebox{3cm}{\columnwidth}{\includegraphics*{h1235.f1b}}}
	\caption{
		  Best fit photon spectrum and contributions to $\chi^2$ to SIS-S0 data
                  of \object{MS 0205.7+3509} using the power law with Galactic absorber
		  and redshifted absorber model.
		}
	\label{spec}
\end{figure}

The deconvolved photon spectra from one of the SIS instruments (SIS-S0) for
the best fit power law with Galactic absorber and redshifted absorber (model
(b)) is plotted in Fig.~\ref{spec}. The slight turnover at the soft end of
the spectrum due to absorption is evident. A formal F-test yields $>99.99\%$
confidence that a power law with Galactic absorber and redshifted absorber
(model (b)) is a better representation of the data than a power law with
Galactic absorber (model (a)).  A broken power law with Galactic absorber
(model (c)), does not yield a better fit despite having one more free
parameter than model (b), and is not considered further.

\begin{table}
	\begin{center}
		\begin{tabular}{l c c c c}
			\hline\hline
			\textbf{Det}
				& \textbf{$N_{H}\times10^{20}$ }
				& \textbf{$\Gamma$}
				& \textbf{$F_{1\mathrm{keV}}\times10^{-3}$}
				& \textbf{$\chi^{2}$/DoF}\\
			
				& ($\mathrm{cm}^{-2}$)
				& \multicolumn{3}{c}{(ph\,keV$^{-1}$cm$^{-2}$s$^{-1}$)}\\
			\hline
			S0 & $21\pm7$ & $2.5\pm0.1$ &  $1.1\pm0.1$ & 83.51/118 \\
			S1 & $27^{+9}_{-8}$ & $2.4^{+0.2}_{-0.1}$ &  $1.1^{+0.2}_{-0.1}$ & 91.32/116 \\
			S2 & $22^{+37}_{-22}$ & $2.8^{+0.4}_{-0.3}$ &  $1.2^{+0.5}_{-0.3}$ & 157.68/167 \\
			S3 & $2^{+28}_{-2}$ & $2.6^{+0.3}_{-0.2}$ &  $1.1^{+0.3}_{-0.1}$ & 168.56/193 \\
			\hline
		\end{tabular}
	\end{center}
	\caption{Best-fit results obtained for model (b) with fixed local absorber
	  $N_{H}=6.1\times10^{20}\mathrm{cm}^{-2}$
	  \protect\cite{1992ApJS...79...77S} and variable absorber at
	  $z=0.318$.  Fit results are shown for each ASCA instrument for
	  comparison and consistency.  Errors are 90\% confidence intervals
	  for 1 parameter of interest.}
	\label{instrumentresults}
\end{table}

\begin{figure}
	\rotatebox{-90}{\resizebox{!}{\columnwidth}{\includegraphics*{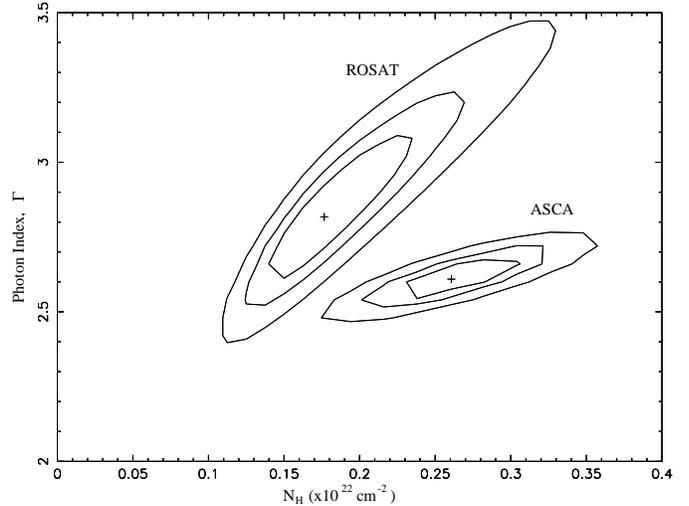}}}
	\caption{A plot of 68\%, 90\% and 99\% confidence contours
		for two parameters of interest, equivalent excess hydrogenic
		absorbing column ($N_{H}$) at a redshift of 0.318, and photon
		index ($\Gamma$) from a simultaneous fit to model (b), of the data
		from all four ASCA instruments.  Confidence contours from the ROSAT
		observation
		\protect\cite{1995ApJ...454...55S} are included for comparison.}
	\label{cont}
\end{figure}

Confidence contours for the best fit model are plotted in Fig.~\ref{cont}
(for the two parameters, equivalent Hydrogen absorbing column ($N_{H}$) and
photon index ($\Gamma$)).

The results of the spectral fitting to models (a) and (b) are presented in
Table~\ref{ascaresults}.

\begin{table}
	\begin{center}
		\begin{tabular}{l c c c c}
			\hline\hline
			
			& \textbf{$\Gamma$}
			& \textbf{$N_{H}\times10^{20}$ }
			& \textbf{$F_{1\mathrm{keV}}\times10^{-3}$}
			& \textbf{$\chi^{2}$/DoF}\\
			&
			& ($\mathrm{cm}^{-2}$)
			& \multicolumn{2}{c}{(ph\,keV$^{-1}$cm$^{-2}$s$^{-1}$)}\\
			\hline
			a & $2.22\pm0.04$ & --- & $0.83\pm0.02$ & 635.72/604\\
			b & $2.61_{-0.08}^{+0.09}$ & $26\pm5$ & $1.20_{-0.08}^{+0.09}$ & 530.78/603\\
			\hline
		\end{tabular}
	\end{center}
	\caption{ASCA best-fit results obtained for models (a) and (b) using
		combined SIS and GIS data.  Both models have a fixed local absorber
		$N_{H}=6.1\times10^{20}\mathrm{cm}^{-2}$
		\protect\cite{1992ApJS...79...77S}.  Errors are 90\%
		confidence intervals for 1 parameter of interest.}
	\label{ascaresults}
\end{table}

Visual inspection of the x--ray lightcurve yields no evidence for
variability greater than 3$\sigma$ on timescales of hours. Optical
photometry on February 5 and 6 yielded a V-band magnitude of
$18.6\pm0.2$. No source variability was detected between the eight exposures
above a level of $0.05\mbox{ mag}$.

\begin{figure}
	\resizebox{\columnwidth}{!}{\includegraphics*{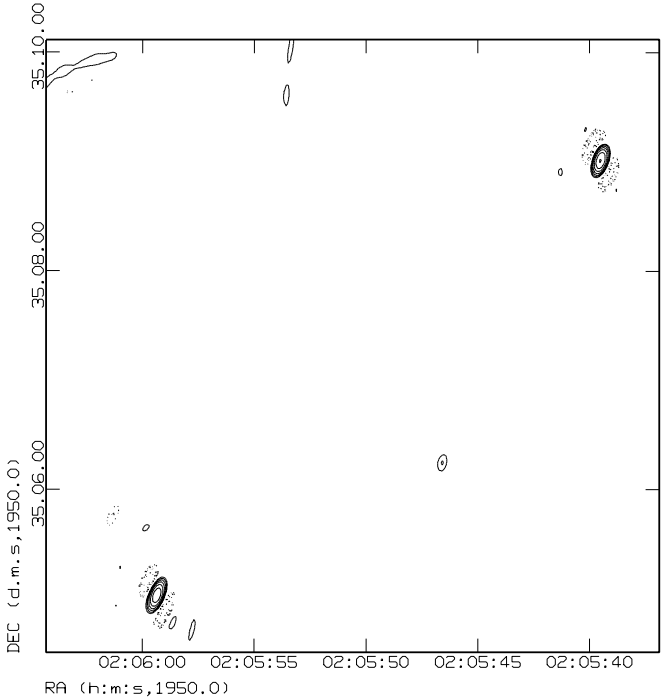}}
	\label{radio_map} 
	\caption{Radio map at 6\,cm of the field surrounding \object{MS 0205.7+3509}
		observed with the Westerbork Synthesis Radio Telescope. 
		\object{MS 0205.7+3509} is at top right
		(RA (1950) = 02h05m39.51s,  Dec (1950) =
		35\degr09\arcmin00.94\arcsec).  The coordinates and fluxes
		of the three sources in the field are given in
		Table~\ref{radiotable}.}
\end{figure}

Radio observations during the period 4--7 February yield a flux value of
$5.75\pm0.125\mbox{ mJy}$ for \object{MS 0205.7+3509}, which is inconsistent
with previous 6\,cm measurements obtained at the VLA (4.91$\pm0.02$\,mJy)
\cite{1995ApJ...454...55S},
 a variability typical of BL Lacs (see for example \cite{1990A&A...235L...1W}).
  Two radio sources were detected close to
\object{MS 0205.7+3509} (Fig.~\ref{radio_map} ) and their fluxes and
positions are given in Table~\ref{radiotable}.

\begin{table}
	\begin{center}
		\begin{tabular}{c c c}
			\hline\hline
			RA (B1950) & Dec (B1950) & Flux (mJy)\\
			\hline
			02:05:39.51 & 35:09:00.94 & 5.75\\
			02:05:46.58 & 35:06:14.83 & 1.02\\
			02:05:59.35 & 35:05:01.75 & 5.29\\
			\hline
	          \end{tabular}
	\end{center}
	\caption{Results of the 6\,cm radio observations carried out at
		WSRT. The two other sources in the field are several arcminutes away from
		\object{MS 0205.7+3509}.}
	\label{radiotable}
\end{table}

As can be seen from Fig.~\ref{cont}, these ASCA results do not agree with
the ROSAT measurement of 1992 \cite{1995ApJ...454...55S}.  When both data
sets are compared over their overlapping energy range, $0.5$--$1.7$\,keV,
the photon index, $\Gamma$ and the equivalent hydrogen column, $N_{H}$ do
not agree to within 2$\sigma$ (two parameters of interest).
Fixing the
redshifted absorption at the ROSAT level, and fitting the ASCA data again 
over the overlapping energy region, an F-test indicates that this fit is
not significantly better than a fit with freely variable absorption.
For this fit the ROSAT $\Gamma=2.8^{+0.3}_{-0.2}$ and ASCA
$\Gamma=2.5\pm0.5$.

%---------------------------DISCUSSION-------------------------------------

\section{Discussion}

Although BL Lacs were originally classified as XBLs or RBLs depending on the
energy band in which they were first discovered, recent work has focused on
determining the physical distinctions between the two types of source
\cite{1995ApJ...444..567P,fcgm97,1998ApJ...506..621G}.  For example, the
spectral energy distributions of XBLs and RBLs differ markedly, with the peak in synchrotron
output occurring in the x--ray range for the former and the radio range for
the latter. Such differences may arise due to changes in the dominant
radiation mechanism occurring as a function of both the angle between the
line of sight and the bulk velocity \cite{mgtt86}. Synchrotron emission is
expected to dominate the x--ray spectra of XBLs (viewed at a large angle)
while inverse compton (IC) emission should dominate the x--ray spectrum of
RBLs (viewed at small angles and significantly Doppler boosted). Another
interpretation of the different properties of XBLs and RBLs is based on the
cut-off frequencies of the synchrotron spectral energy distribution
\cite{1995ApJ...444..567P}. In this model BL Lacs with a cut-off in the
IR/optical band are RBLs, while those with a cut-off in the UV/x--ray region
are XBLs.
  
Empirically, a source may be classified as an XBL if it has
$\log(S_{x}/S_{r})\geq-5.5$, where $\mathrm{S}_{x}$ is the flux at 2\,keV
and $\mathrm{S}_{r}$ is the flux at 5\,GHz, defined in the same units
\cite{1997astro.ph..6102K}. \object{MS 0205.7+3509} lies firmly in the XBL category,
with $\log(S_{x}/S_{r})=-4.17$ derived from the simultaneous observations
presented here, suggesting that synchrotron emission is dominating the
x--ray emission in the ASCA energy band.

The shape of the spectrum from optical to x--ray wavelengths is an
indicator of the relative importance of the IC component to the
continuum emission \cite{1996ApJ...463..444S}. 
The parameter  $\alpha_{xox} = \alpha_{ox}-\alpha_{x}$,   
where $\alpha_{x}$ is the ROSAT PSPC energy index and
\begin{eqnarray}
    \alpha_{ox}=-\frac{\log(F_{o}/F_{x})}{\log(\nu_{o}/\nu_{x})}\nonumber 
\end{eqnarray}
is a measure of the concavity/convexity of the spectrum between
optical and x--ray wavelengths. A positive value for $\alpha_{xox}$
(concave shape) implies the presence of a hard IC component.
The following  values  are obtained for \object{MS 0205.7+3509}:
$\alpha_{xox}=-0.82$, $\alpha_{ox}=0.79$, $\alpha_{ro}=0.32$ and
$\alpha_{rx}=0.48$.
The simultaneous spectrum therefore shows no evidence of
concavity  out to $\sim 10$\,keV.
On the basis of these data an IC component may be expected
to become significant in the hard x--ray region (clearly above 10\,keV) and
this could be verified by spectroscopic observations in this energy band
\cite{1996ApJ...463..444S}.  Three nearby XBLs (\object{Mkn 421}, \object{Mkn 501} and
\object{1ES 2344+514}) have been detected at TeV energies and this high energy
emission may be attributable to IC scattering of the highest energy
electrons off the synchrotron x--ray photons \cite{wafk97,cbb+97}. 
Calculations of the opacity of the universal soft photon background to
$\gamma$--rays suggest that \object{MS 0205.7+3509} may be detectable above 100\,GeV
by future $\gamma$--ray telescopes \cite{ss98}.  We note that
\object{MS 0205.7+3509} (like \object{Mkn 501}) was not detected by EGRET
with a 2$\sigma$ upper limit of $0.9\times 10^{-7}$ ph/cm$^2$/s at E$>$100MeV
\cite{fbc+94}.

With $\log(S_{x}/S_{r})=-4.17$ and $\alpha_{xox}=-0.82$,
\object{MS0205.7+3509} emerges as a typical XBL. It lies at the
furthest extent of the XBL category on the $\alpha_{ro}$ vs. $\alpha_{xox}$
diagram and near the middle of the XBL category in the $\log(S_{x}/S_{r})$ diagram.
The convexity of the simultaneous
multiwavelength spectrum indicates that the source is not a flat-spectrum
radio quasar (FSRQ) which is being affected by microlensing as suggested by
Stocke et al. \cite*{1995ApJ...454...55S} since it does not have the
characteristic FSRQ concave spectral shape \cite{1996ApJ...463..444S}. 
Furthermore on the basis of the x--ray spectrum only, we can rule out the
possibility that \object{MS 0205.7+3509} is an FSRQ since the x--ray power law
photon indices of that class of AGN are significantly harder than those of
XBLs, lying in the range 1.3--1.9 \cite{1996ApJ...463..424U,1998ApJ...504..693K}.

The ASCA spectrum indicates variability in the x-ray flux of this object,
compared to the spectrum taken with ROSAT in 1992
($F_{1\mathrm{keV}}$ is $1.20_{-0.03}^{+0.03}\times10^{-12}$
erg\,keV$^{-1}$cm$^{-2}$s$^{-1}$ for the ASCA observation of February 1997,
compared to $1.49\pm0.05$ for the ROSAT 1992 observation).  We detect a change in
either the photon index, $\Gamma$ or the equivalent Hydrogen column,
$N_{H}$. Assuming a change in the photon index, $\Gamma$, then it is
interesting to note that this behaviour is quite similar to that observed by
Madejski et al. \cite*{1996ApJ...459..156M} in \object{AO 0235+164} between
their non-simultaneous ROSAT and ASCA observations.

The case for microlensing effects in this source (decentered nucleus from
host, spiral host and the presence of excess soft x-ray absorption) has been
questioned because results from a deep subarcsecond optical study have shown
that there is a companion galaxy $2''$ away from the BL Lac. When it is
removed, the BL Lac is centred with respect to the remaining nebulosity and
the properties of the host galaxy do not look anomalous
\cite{1997AA...321..374F}. However, the existence of excess x--ray
absorption in the ROSAT PSPC data is also clearly detected in the ASCA
spectrum and may be attributable to the halo of this companion galaxy, if it
is in a foreground location. Another possibility is that the BL Lac resides
behind its nominal `host' galaxy.

%----------------------------CONCLUSIONS-------------------------------------

\section{Conclusions}

Simultaneous multiwavelength observations of \object{MS 0205.7+3509}
demonstrate that it is a typical XBL, with $\alpha_{xox}=-0.82$.
No evidence for an IC component was observed in the ASCA spectrum and
on the basis of these results such a component is expected to appear in
the hard x--ray ($>10$\,keV) band. The ASCA spectrum confirms a soft
x--ray absorption in excess of the Galactic value which was previously
identified with ROSAT.
However, the location and nature of this foreground absorber and its
possible role in microlensing remains unclear. More extensive optical
monitoring could reveal rapid variability indicating probable gravitational
microlensing in \object{MS 0205.7+3509} as has been done in the case of
\object{AO 0235+164} \cite{1996A&A...310....1R}.
 
\bibliographystyle{h1235}
\bibliography{h1235}

\end{document}